\documentclass[conference]{IEEEtran}
\usepackage{cite}
\usepackage{amsmath,amssymb,amsfonts}
\usepackage{graphicx}
\usepackage{textcomp}
\usepackage{xcolor}
\usepackage{hyperref}
\usepackage{algorithm}
\usepackage{algpseudocode}
\usepackage{enumitem}
\usepackage{float}
\usepackage{subcaption}
\begin{document}

\title{3D-PNAS: 3D Industrial Surface Anomaly Synthesis with Perlin Noise}

\author{\IEEEauthorblockN{Yifeng Cheng\IEEEauthorrefmark{1} and Juan Du\IEEEauthorrefmark{1}}
\IEEEauthorblockA{\IEEEauthorrefmark{1}The Hong Kong University of Science and Technology (Guangzhou)\\
Guangzhou, China}
}

\maketitle
\begin{abstract}
    Large pretrained vision foundation models have shown significant potential in various vision tasks. However, for industrial anomaly detection, the scarcity of real defect samples poses a critical challenge in leveraging these models. While 2D anomaly generation has significantly advanced with established generative models, the adoption of 3D sensors in industrial manufacturing has made leveraging 3D data for surface quality inspection an emerging trend. In contrast to 2D techniques, 3D anomaly generation remains largely unexplored, limiting the potential of 3D data in industrial quality inspection. To address this gap, we propose a novel yet simple 3D anomaly generation method, 3D-PNAS, based on Perlin noise and surface parameterization. Our method generates realistic 3D surface anomalies by projecting the point cloud onto a 2D plane, sampling multi-scale noise values from a Perlin noise field, and perturbing the point cloud along its normal direction. Through comprehensive visualization experiments, we demonstrate how key parameters—including noise scale, perturbation strength, and octaves—provide fine-grained control over the generated anomalies, enabling the creation of diverse defect patterns from pronounced deformations to subtle surface variations. Additionally, our cross-category experiments show that the method produces consistent yet geometrically plausible anomalies across different object types, adapting to their specific surface characteristics. We also provide a comprehensive codebase and visualization toolkit to facilitate future research.
\end{abstract}
\begin{IEEEkeywords}
3D anomaly synthesis, industrial inspection, point clouds, anomaly detection, Perlin noise
\end{IEEEkeywords}

\section{Introduction}
Industrial quality control increasingly relies on automated inspection systems. However, defective samples are scarce in reality \cite{liuDeepIndustrialImage2024,xuSurveyIndustrialAnomalies2025}. With the development of 2D methods \cite{liuDeepIndustrialImage2024} and well-established generative models like GANs and Diffusion models, 2D synthesis methods have become advanced, capable of generating diverse, controllable, and realistic anomalies \cite{Ali2024AnomalyControl,Hu2024AnomalyXFusion,huAnomalyDiffusionFewShotAnomaly2024,jiangCAGENControllableAnomaly2024, Jin2024DualAnoDiff,sunCUTControllableUniversal2024,xuSurveyIndustrialAnomalies2025}. High-quality synthesized data is crucial as it bridges the gap when using pre-trained models for industrial anomaly detection and determines the performance of the detection system \cite{xuSurveyIndustrialAnomalies2025}.

A recent trend in industrial anomaly detection is the use of pre-trained foundational vision-language models (VLMs). These models, trained on billions of natural images and their associated text, demonstrate strong generalization capabilities across a wide range of visual tasks. However, a significant domain gap exists when applying these models to industrial anomaly detection\cite{anomalyclip23,winclip23,zuoCLIP3DADExtendingCLIP2024,zhou2024pointad}. This domain gap arises because VLMs like CLIP are primarily trained on large-scale, natural image datasets that differ substantially from industrial data in both content and distribution. Industrial anomaly detection often requires fine-grained recognition of subtle, low-level defects on highly specialized objects, whereas VLMs are optimized for high-level semantic understanding and open-vocabulary classification in natural scenes. As a result, VLMs may lack sensitivity to the local, pixel-level or geometric irregularities that characterize industrial defects, and their representations may not align well with the specific visual cues needed for anomaly detection in manufacturing settings. Existing studies\cite{anomalyclip23,winclip23,zuoCLIP3DADExtendingCLIP2024,zhou2024pointad} have shown promising results with vision-language models like CLIP\cite{clip2021}, but in few-shot settings, labeled anomaly samples are still necessary to adapt these models to the industrial domain. To address this challenge, current approaches either leverage auxiliary data\cite{anomalyclip23,zhou2024pointad} or generate synthesized data\cite{zuoCLIP3DADExtendingCLIP2024} to bridge the gap between the pre-trained model's domain knowledge and the requirements of industrial anomaly detection.

In contrast to well-established 2D industrial anomaly synthesis (IAS) methods \cite{xuSurveyIndustrialAnomalies2025}, 3D IAS is still underexplored. While diffusion models and GANs have demonstrated remarkable success in generating realistic anomalies in 2D images, their direct adaptation to 3D point clouds remains challenging due to the sparse and irregular nature of point cloud data. Although approaches like CutPaste\cite{liCutPasteSelfSupervisedLearning2021} can be adapted to structured point clouds, they often fail to maintain geometric plausibility, producing anomalies that do not conform to the physical characteristics of real industrial defects. Additionally, the diversity of synthesized anomalies is limited, making it difficult to cover the complex types of defects encountered in industrial settings.

\begin{figure*}[t]
    \centering
    \begin{subfigure}[b]{0.49\linewidth}
        \centering
        \includegraphics[width=\linewidth]{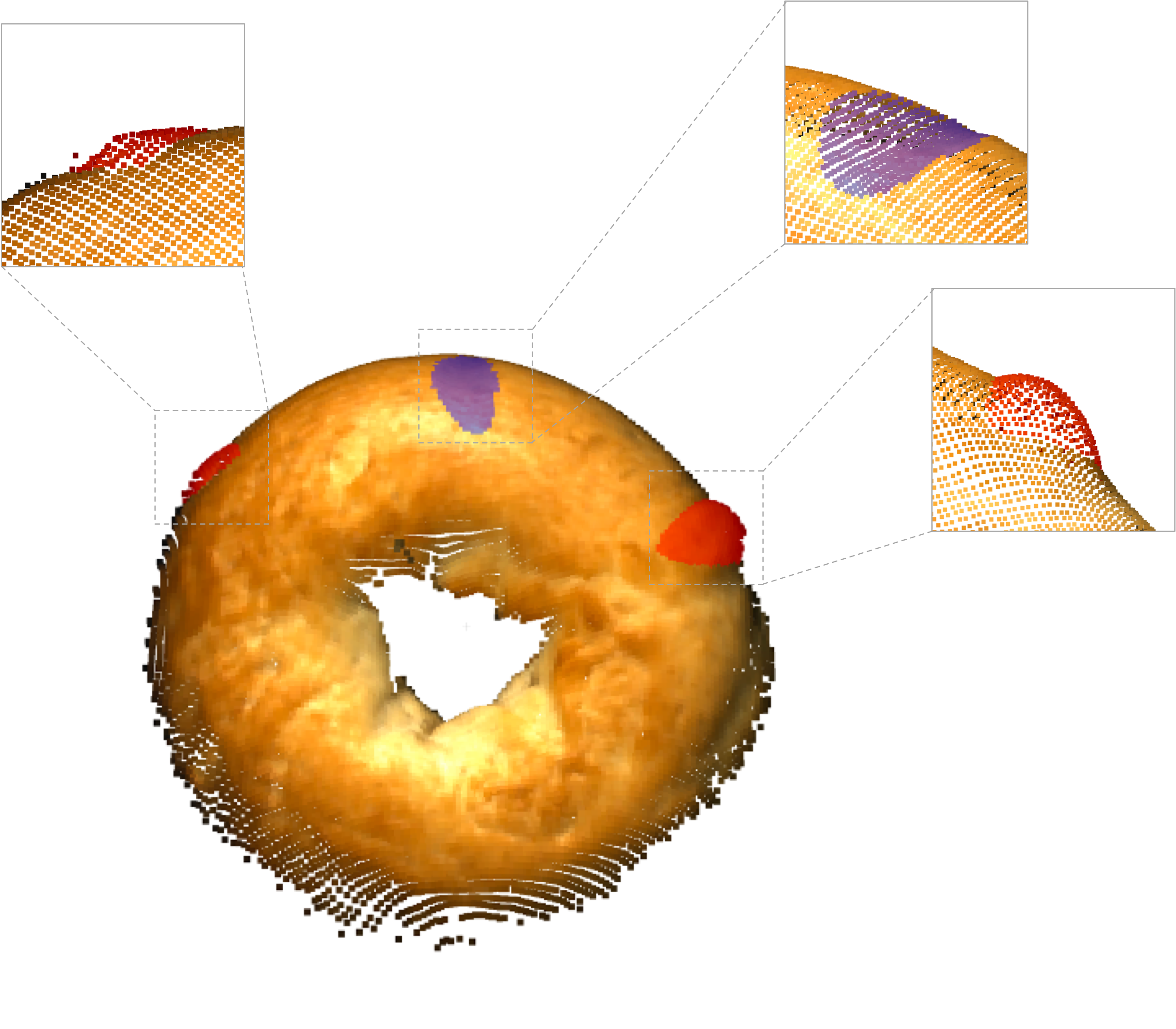}
        \caption{Our approach maintains surface continuity and considers local geometry, producing more realistic industrial defects corresponding to the underlying structure of the object.}
        \label{fig:cutpaste_a}
    \end{subfigure}
    \hfill
    \begin{subfigure}[b]{0.4\linewidth}
        \centering
        \includegraphics[width=\linewidth]{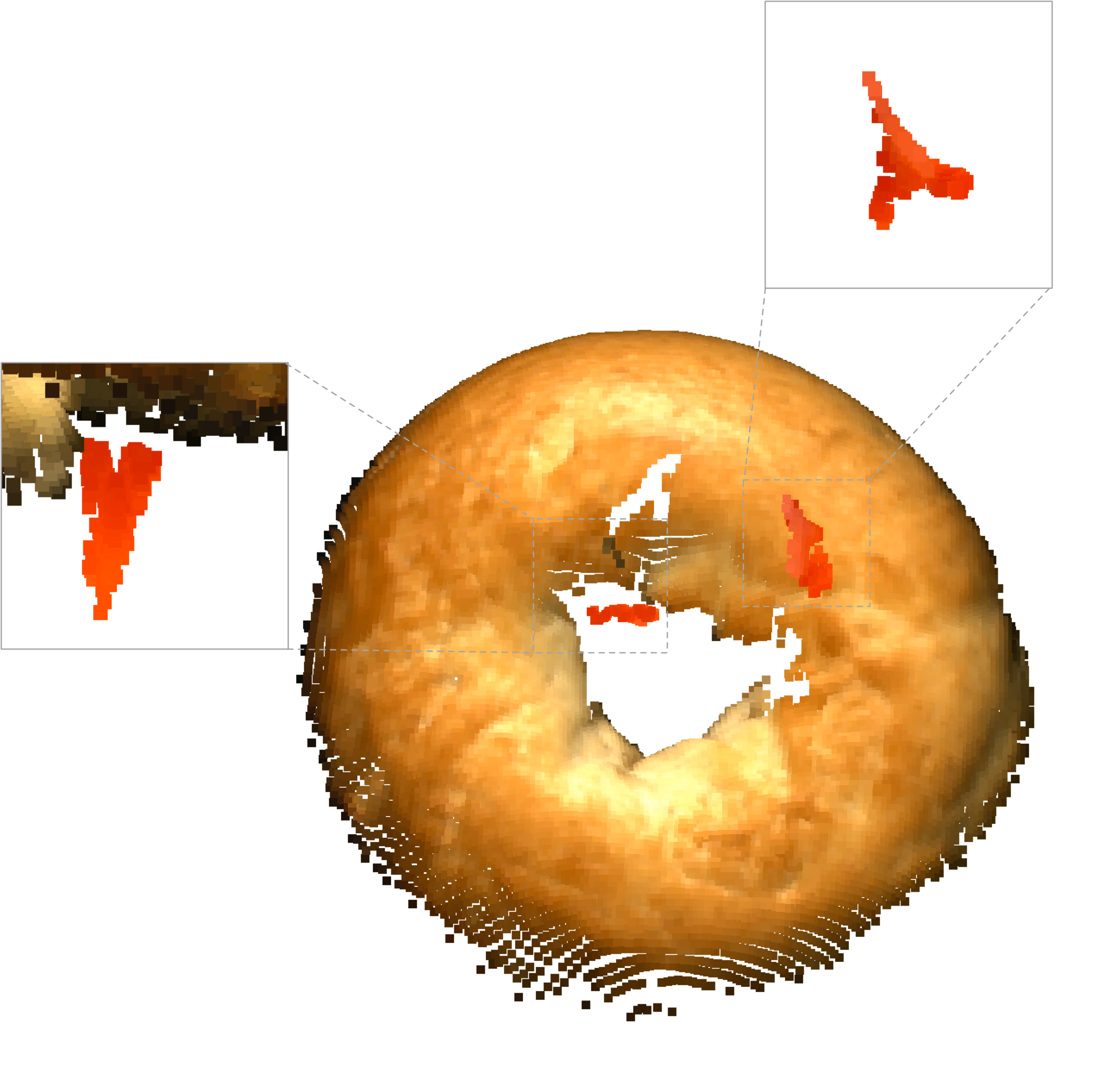}
        \caption{ An adaptation  of cut-paste\cite{liCutPasteSelfSupervisedLearning2021} method to 3D \cite{taog2miad2025} fails to preserve geometric consistency, resulting in physically implausible anomalies.}
        \label{fig:cutpaste_b}
    \end{subfigure}
    \caption{Comparison between cut-paste based approach and our approach in 3D anomaly synthesis.}
    \label{fig:cutpaste_comparison}
\end{figure*}

In recent years, with the wide adoption of 3D sensors in surface inspection, effectively leveraging 3D data has become crucial\cite{du3DVisionbasedAnomaly2025a}.
To effectively leverage 3D data and adapt pre-trained foundational models for multi-modal industrial anomaly detection, high-quality synthesized data becomes increasingly important. While several works have explored 3D anomaly synthesis methods\cite{zavrtanikKeepDRAEMingDiscriminative2024,liDAS3DDualmodalityAnomaly2024,yePO3ADPredictingPoint2024,zhouR3DADReconstructionDiffusion2025}, there remains a notable absence of widely adopted 3D anomaly generation techniques. To address these limitations, we propose a simple yet effective 3D anomaly synthesis approach, accompanied by a comprehensive codebase and visualization toolkit to facilitate the application of these algorithms in enhancing 3D anomaly detection.

Our contributions are summarized as follows:
\begin{itemize}
    \item We propose a simple yet effective 3D anomaly generator that synthesizes training data with physically plausible anomalies using  Perlin noise.
    \item We provide a comprehensive codebase and visualization toolkit to use of our method in 3D anomaly generation, to enhance future research on 3D anomaly detection.
    \item We also conduct visualization studies to demonstrate the  parameters of our algorithm result in change of generated 3D anomalies, showcasing the capability of our algorithms to generate diverse anomalies.
\end{itemize}


\section{Related Work} 

\subsection{2D Anomaly Synthesis} 

2D anomaly generation can be categorized into four broad approaches\cite{xuSurveyIndustrialAnomalies2025}. Hand-craft approaches rely on manually designed rules, such as cropping, pasting, or inpainting, to simulate anomalies in a simple and efficient manner, though with limited diversity \cite{liCutPasteSelfSupervisedLearning2021,nsa2022,zavrtanikReconstructionInpaintingVisual2021}.
Hypothesis-based approaches model the statistical distribution of normal data and generate anomalies by perturbing features in the latent space, offering controlled but feature-level anomalies \cite{chenProgressiveBoundaryGuided2025,Liu2023SimpleNet}.
Generative model-based approaches leverage advanced models like GANs and diffusion models to synthesize realistic anomalies, either through full-image generation, image to image translation, or local replacement, achieving high quality results with the cost of computational complexity \cite{zhangDefectGANHighFidelityDefect2021,zhangRealNetFeatureSelection2024}.
VLM-based approaches utilize large-scale vision-language models and multimodal input (e.g., text, masks) to generate context-aware anomalies with precise control over type, location, and intensity, while requiring significant computational resources and available multimodal data \cite{jiangCAGENControllableAnomaly2024, sunCUTControllableUniversal2024}.

\subsection{3D Anomaly Synthesis}  

3D anomaly synthesis is a nascent but evolving field. Unlike 2D anomaly synthesis, which have various school of methods that can fall in a well-established taxonomy (hand-crafted, hypothesis-based, generative, and VLM-based)\cite{xuSurveyIndustrialAnomalies2025}, 3D methods remain largely exploratory and majority of 3D anomaly synthesis methods fall into the hand-crafted category observed in 2D, characterized by manually designed rules or operations to simulate anomalies.

Methods developed for 3D anomaly synthesis can be categorized based on the type of point cloud they target. Some methods focus on structured 2.5D point clouds, which are captured by depth cameras and result in a 2D image-like structure like MVTec-3D AD dataset \cite{bergmannMVTec3DADDataset2022}. For example, DAS3D \cite{liDAS3DDualmodalityAnomaly2024} generates depth anomalies using Perlin noise to create trinary masks (-1, 0, 1), which are smoothed with a skewed Gaussian filter to simulate realistic depth changes, then added to the original depth image. Keep DRÆMing \cite{zavrtanikKeepDRAEMingDiscriminative2024} also employs Perlin noise with thresholding to obtain mask, but directly uses Perlin noise value and scaling the value to generate smooth depth values later added to the original depth image to create anomalous region. Similarly, Tao et al. \cite{taog2miad2025} using Perlin noise to generate binarized masks for anomaly placement in their G$^2$SF-MIAD framework. Source regions from normal samples are corrupted through random augmentations—pixel modifications for RGB images and geometric translations for 3D point clouds—before being pasted onto target data. 

In contrast, other methods target unstructured 3D point clouds, typically derived from 3D meshes. Unlike methods that work with structured 2.5D data, they cannot directly apply deformations to point clouds by adding depth values. These methods address the challenges of unstructured point clouds, by incorporating geometric information like normal vectors or use viewpoint-based patch selection. For instance, PO3AD \cite{yePO3ADPredictingPoint2024} shifts points in selected patches along normal vectors, with the shift distance inversely proportional to the distance from the patch centroid. R3D-AD \cite{zhouR3DADReconstructionDiffusion2025} introduces the Patch-Gen approach, which selects patches by randomly sampling viewpoints on a surrounding sphere and applies controlled translations to form damage, sinks, and bulges. Instead of creating anomalies algorithmically, Anomaly-ShapeNet \cite{anomlyshapenet} manually carves defects such as bulges, concavities, and cracks onto ShapeNet meshes using Blender, an open-source 3D modeling software.

\subsection{Research Gaps and Opportunities}
An emerging trend is the use of Vision Foundation Models (VFMs) for multi-modal few-shot/zero-shot anomaly generation. Recent works, such as CLIP-3D-AD \cite{zuoCLIP3DADExtendingCLIP2024} and PointAD \cite{zhou2024pointad}, leverage CLIP for 3D anomaly detection by projecting 3D data into 2D views. However, these methods rely on 2D representations rather than directly utilizing 3D backbones, highlighting the need for synthetic anomaly data to bridge this gap.

In industrial settings, 3D data is often collected as 2.5D point clouds, representing partial surfaces. While methods like DAS3D \cite{liDAS3DDualmodalityAnomaly2024}, Keep DRÆMing \cite{zavrtanikKeepDRAEMingDiscriminative2024}, and Tao et al. \cite{taog2miad2025} all rely on structured 2.5D point clouds, they do not fully exploit surface normal information, which is crucial for generating physically plausible anomalies. Although these methods can effectively generate synthetic anomalies for 2.5D point clouds in industrial settings, they do not incorporate surface normal information enhance the physical plausibility of anomalies. Our approach, by contrast, assumes only planar surfaces, do not  explicitly assume that the point clouds are structured, offering greater flexibility and adaptability to diverse industrial scenarios, addressing this gap in generating more realistic 3D anomalies. 


\section{Methodology}
\begin{figure*}[t]
    \centering
    \includegraphics[width=0.9\linewidth]{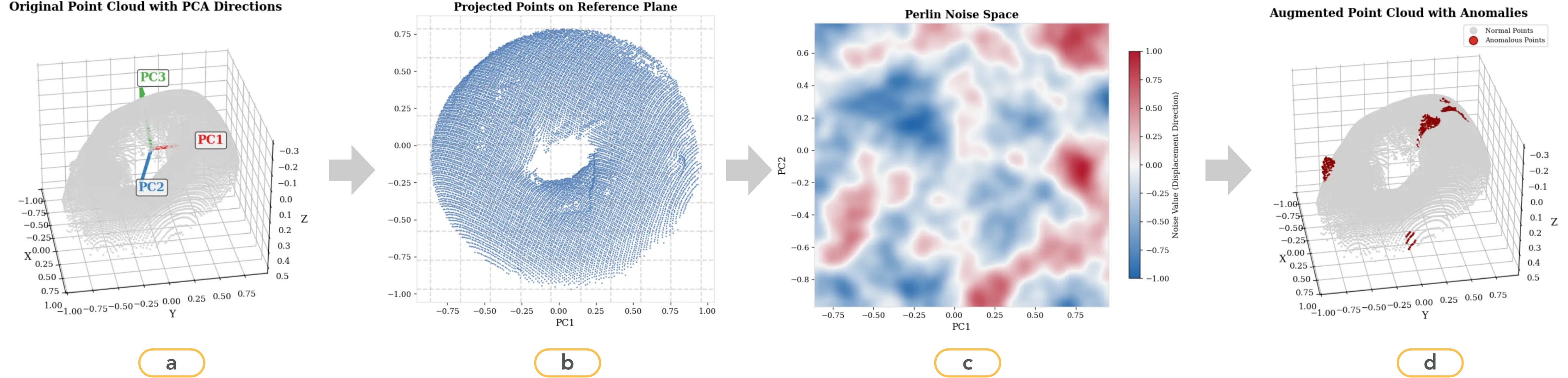}
    \caption{Overview of our 3D anomaly synthesis method: (a) Original point cloud with PCA directions; (b) Projected points on reference plane; (c) Perlin noise space; and (d) Augmented point cloud with anomalies. This pipeline ensures that the generated anomalies respect the local surface geometry while providing diverse and controllable deformations.}
    \label{fig:overview}
\end{figure*}

Our goal is to introduce controlled \emph{surface-based} deformations (anomalies) on a 3D point cloud using Perlin noise. The overall pipeline can be divided into four main stages: (i) surface parameterization, (ii) noise generation on the 2D reference plane, (iii) computation of surface normals, and (iv) threshold-based anomaly application with adaptive masking. Figure \ref{fig:overview} illustrates the complete pipeline of our proposed method.

\subsection{Surface Parameterization}
Given a point cloud $\mathbf{P} \in \mathbb{R}^{N_p\times 3}$ where $N_p$ indicates the total number of 3D points, we first discard invalid points (e.g., the points that depth sensor does not return a valid depth value, or the points that are background). Denote the remaining valid points by $\mathbf{P}_{\mathrm{valid}} \in \mathbb{R}^{M\times 3}$, where $M \leq N_p$. We then compute a 2D parameterization of $\mathbf{P}_{\mathrm{valid}}$ onto a plane via PCA:

\begin{enumerate}
    \item \emph{Centering.} Compute the centroid:
    \[
      \boldsymbol{\mu} \;=\; \frac{1}{M}\,\sum_{i=1}^M \mathbf{p}_i,
    \quad \mathbf{p}_i \in \mathbf{P}_{\mathrm{valid}},
    \]
    and let 
    \[
      \mathbf{P}_c \;=\; \mathbf{P}_{\mathrm{valid}} \;-\; \boldsymbol{\mu}.
    \]
    \item \emph{PCA via SVD.} Perform singular value decomposition (SVD) on $\mathbf{P}_c$:
    \[
      \mathbf{P}_c \;=\; \mathbf{U}\,\mathbf{S}\,\mathbf{V}^\mathsf{T},
    \]
    where $\mathbf{V}^\mathsf{T}\in \mathbb{R}^{3\times 3}$ contains the principal directions.
    \item \emph{Projection to 2D.} Let $\mathbf{M} = \mathbf{V}^\mathsf{T}_{1:2}$ (the top two principal directions, forming a $3\times 2$ matrix). The 2D projection is 
    \[
      \mathbf{P}_{2D} \;=\; \mathbf{P}_c\,\mathbf{M} \;\in\; \mathbb{R}^{M \times 2}.
    \]
\end{enumerate}

\noindent
Thus, each valid 3D point $\mathbf{p}_i$ maps to a 2D coordinate $\mathbf{p}_{2D,i}$, easing subsequent noise generation and interpolation.

\subsection{Continuous Perlin Noise Generation}
We generate a continuous 2D Perlin noise field over a bounding box that covers all $\mathbf{P}_{2D}$.

\begin{enumerate}
    \item \emph{2D Bounds.} Let
    \[
      \boldsymbol{\beta}_{\min} \;=\; \min(\mathbf{P}_{2D}), 
      \quad
      \boldsymbol{\beta}_{\max} \;=\; \max(\mathbf{P}_{2D}),
    \]
    where the minima and maxima are taken elementwise. We then form a regular 2D grid $\mathbf{G} \in \mathbb{R}^{r\times r \times 2}$ with resolution $r \in \mathbb{Z}^+$, whose $xy$-values span $[\boldsymbol{\beta}_{\min}, \boldsymbol{\beta}_{\max}]$. 
    \item \emph{Noise Field.} For each grid point $\mathbf{g}_{ij}$, we compute a Perlin noise value
    \[
      P_{ij} \;=\; \mathrm{Perlin} \Bigl(\mathbf{g}_{ij}\cdot s,\; o,\; p,\; l \Bigr),
    \]
    where $s$, $o$, $p$, $l$ denote noise-scale, octaves, persistence, and lacunarity, respectively. We then min-max normalize
    \[
      \tilde{P}_{ij} \;=\; \frac{\,P_{ij}\;-\;\min_{i,j}P_{ij}\,}{\max_{i,j}P_{ij}\;-\;\min_{i,j}P_{ij}},
      \quad
      \tilde{P}_{ij} \in [0,1].
    \]
    Finally, we shift and scale to $[-1,1]$:
    \[
      P_{ij}^{(-1,1)} \;=\; 2\,\tilde{P}_{ij} \;-\; 1.
    \]
\end{enumerate}

\noindent
Each valid point $\mathbf{p}_{2D,i}$ in $\mathbf{P}_{2D}$ obtains its noise value $\nu_i$ via bilinear interpolation of $P_{ij}^{(-1,1)}$ in the grid.

\subsection{Surface Normal Computation}
We compute surface normals $\mathbf{n}_i$ for each valid 3D point $\mathbf{p}_i$. Let $k \in \mathbb{Z}^+$ be the neighborhood size (e.g., $10$). For each point $\mathbf{p}_i$, we select its $k$ nearest neighbors $\{\mathbf{p}_{j}\}_{j \in \mathcal{N}_i}$ based on Euclidean distance, where $\mathcal{N}_i$ represents the index set of the $k$ neighbors of $\mathbf{p}_i$. We center these neighbors and form a matrix:
\[
  \mathbf{Q}_c = \begin{bmatrix} 
    \mathbf{p}_{j} - \overline{\mathbf{p}}_{\mathcal{N}_i} 
  \end{bmatrix}_{j \in \mathcal{N}_i} \in \mathbb{R}^{|\mathcal{N}_i| \times 3},
\]
where $\overline{\mathbf{p}}_{\mathcal{N}_i}$ is the mean of the neighbors of $\mathbf{p}_i$, defined as:
\[
  \overline{\mathbf{p}}_{\mathcal{N}_i} = \frac{1}{|\mathcal{N}_i|} \sum_{j \in \mathcal{N}_i} \mathbf{p}_j.
\]
We compute the covariance matrix
\[
  \mathbf{C} \;=\; \mathbf{Q}_c^\mathsf{T} \,\mathbf{Q}_c,
\]
and solve the eigenvalue problem
\[
  \mathbf{C}\,\mathbf{v} \;=\; \lambda\,\mathbf{v}.
\]
The eigenvector corresponding to the smallest eigenvalue $\lambda_{\min}$ is taken as the local normal $\mathbf{n}_i$; we additionally enforce outward orientation by ensuring $\mathbf{n}_i$ points consistently relative to its local neighborhood.

\subsection{Threshold-Based Anomaly Application}
We now combine the interpolated noise values $\nu_i$ with normals $\mathbf{n}_i$ to displace each valid point $\mathbf{p}_i$. Let $\alpha \in \mathbb{R}^+$ be the desired \emph{perturbation strength}, and let $\tau \in (0, 1)$ be an initial threshold. We form a binary mask
\[
  \mathbf{m}_i 
  \;=\; 
  \begin{cases}
      1, & \text{if }\, |\nu_i| > \tau,
      \\
      0, & \text{otherwise}.
  \end{cases}
\]
Let $\mathrm{avgMask} = \frac{1}{M}\sum_{i=1}^M m_i$ be the ratio of points exceeding the threshold. If $\mathrm{avgMask} > \rho$ (where $\rho \in (0, 1)$ is the target mask ratio), we \emph{self-adjust} $\tau$ by sorting $\{\lvert \nu_i\rvert \}$ and selecting the value $\tau_{\mathrm{new}}$ that yields $\mathrm{avgMask} = \rho$. This updated threshold is used to refine $\mathbf{m}_i$. 

\paragraph{Local Normalization.} For all masked points $i$ where $m_i = 1$, we preserve the sign of $\nu_i$ but re-scale its magnitude to  $[0,1]$ for $|\nu_i| \ge \tau$. Specifically,
\[
  \widehat{\nu}_i 
  \;=\;
  \begin{cases}
     \mathrm{sign}(\nu_i)\,\dfrac{\;\lvert \nu_i\rvert - \tau\,}{\,1 - \tau\,}, & m_i=1,
     \\
     0, & m_i=0.
  \end{cases}
\]
Finally, the displacement of point $\mathbf{p}_i$ is:
\[
  \boldsymbol{\delta}_i \;=\;\widehat{\nu}_i \,\alpha \;\mathbf{n}_i \,m_i,
\]
and the \emph{augmented} coordinates become
\[
  \mathbf{p}_i^{(\mathrm{aug})} \;=\;\mathbf{p}_i \;+\;\boldsymbol{\delta}_i.
\]
Once displacements are computed for all valid points, we reintegrate any previously invalid points (unchanged, as zero coordinates) and, if needed, reshape back to an original organized structure (e.g., $H\times W\times 3$). This yields the final \emph{anomalous} point cloud $\mathbf{P}_{\mathrm{aug}}$.

\subsection{Rationale}
The crucial insight of our approach is modeling deformations \emph{directly on the surface}. By parameterizing onto a plane (using PCA) and mapping a continuous Perlin noise field back to 3D, we ensure that (i) anomalies respect the local surface structure, and (ii) noise-driven displacements are smoothly varying. The threshold adaptation step ensures a prescribed fraction of points undergo deformation, while \emph{local normalization} of noise values ensuring smooth boundary of generated anomaly region.


\section{Visualization}
In order to demonstrate that our synthetic anomaly generation algorithm is capable of generating realistic synthetic anomaly samples, we conducted a series of visualization experiments on the MV-Tec 3D-AD dataset. We present the results of our experiments across three key dimensions: parameter exploration, anomaly type categorization, and cross-category consistency.

\subsection{Parameter Exploration}
To demonstrate the flexibility and control offered by our method, we systematically explored the parameter space of our Perlin noise-based anomaly generator. We varied five key parameters that control different aspects of the generated anomalies:

\begin{itemize}
    \item \textbf{Noise scale ($s$)}: Controls the frequency of the Perlin noise pattern, with lower values producing larger, smoother anomalies and higher values creating smaller, more frequent anomalies. We tested values of $s \in \{1.0, 2.0, 4.0\}$.
    
    \item \textbf{Perturbation strength ($\alpha$)}: Determines the magnitude of the deformation along the surface normal, with higher values creating more pronounced deformations. We tested values of $\alpha \in \{0.01, 0.02, 0.05\}$.
    
    \item \textbf{Octaves ($o$)}: Determines the level of detail in the noise pattern, with higher values adding finer details. We tested values of $o \in \{1, 2, 4\}$.

\end{itemize}

\begin{figure*}[ht]
    \centering
    \includegraphics[width=0.8\linewidth]{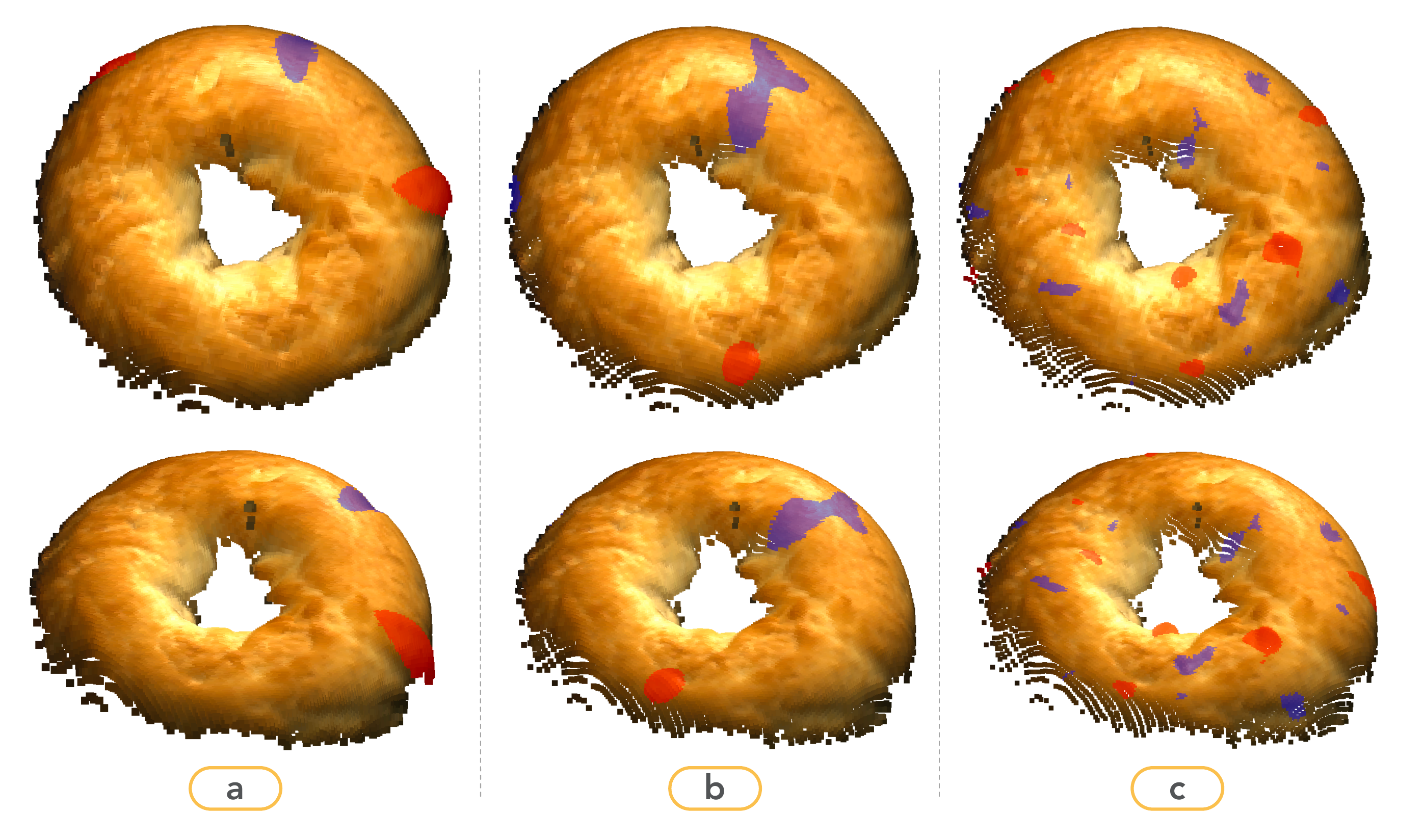}
    \caption{Visualization of the effect of varying the noise scale ($s$) on the generated anomalies in a 3D point cloud: (a) $s=1.0$ produces large, smooth, and sparse deformations; (b) $s=2.0$ creates medium-sized deformations; and (c) $s=4.0$ results in smaller, more localized, and densely distributed anomalies. Red points indicate protrusions, while blue points indicate intrusions.}
    \label{fig:noise_scale}
\end{figure*}

\begin{figure*}[htbp]
    \centering
    \includegraphics[width=0.8\linewidth]{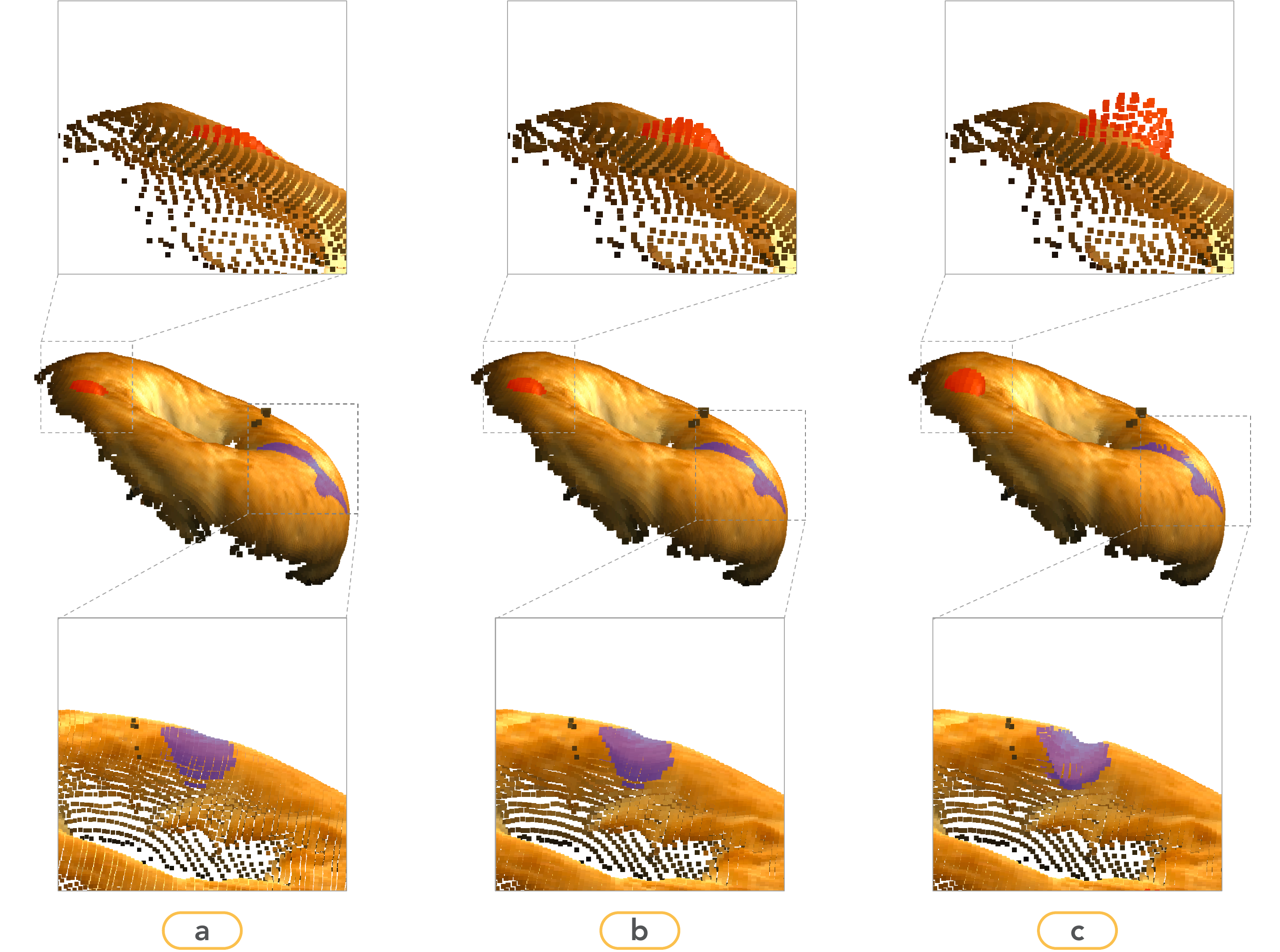}
    \caption{Visualization of the effect of varying the perturbation strength ($\alpha$) on the generated anomalies in a 3D point cloud: (a) $\alpha=0.01$ creates subtle surface modifications; (b) $\alpha=0.02$ produces moderate deformations; and (c) $\alpha=0.05$ results in more pronounced and significant protrusions and intrusions. All cases maintain the same spatial distribution pattern while varying in magnitude.}
    \label{fig:perturb_strength}
\end{figure*}

\begin{figure*}[ht]
    \centering
    \includegraphics[width=0.8\linewidth]{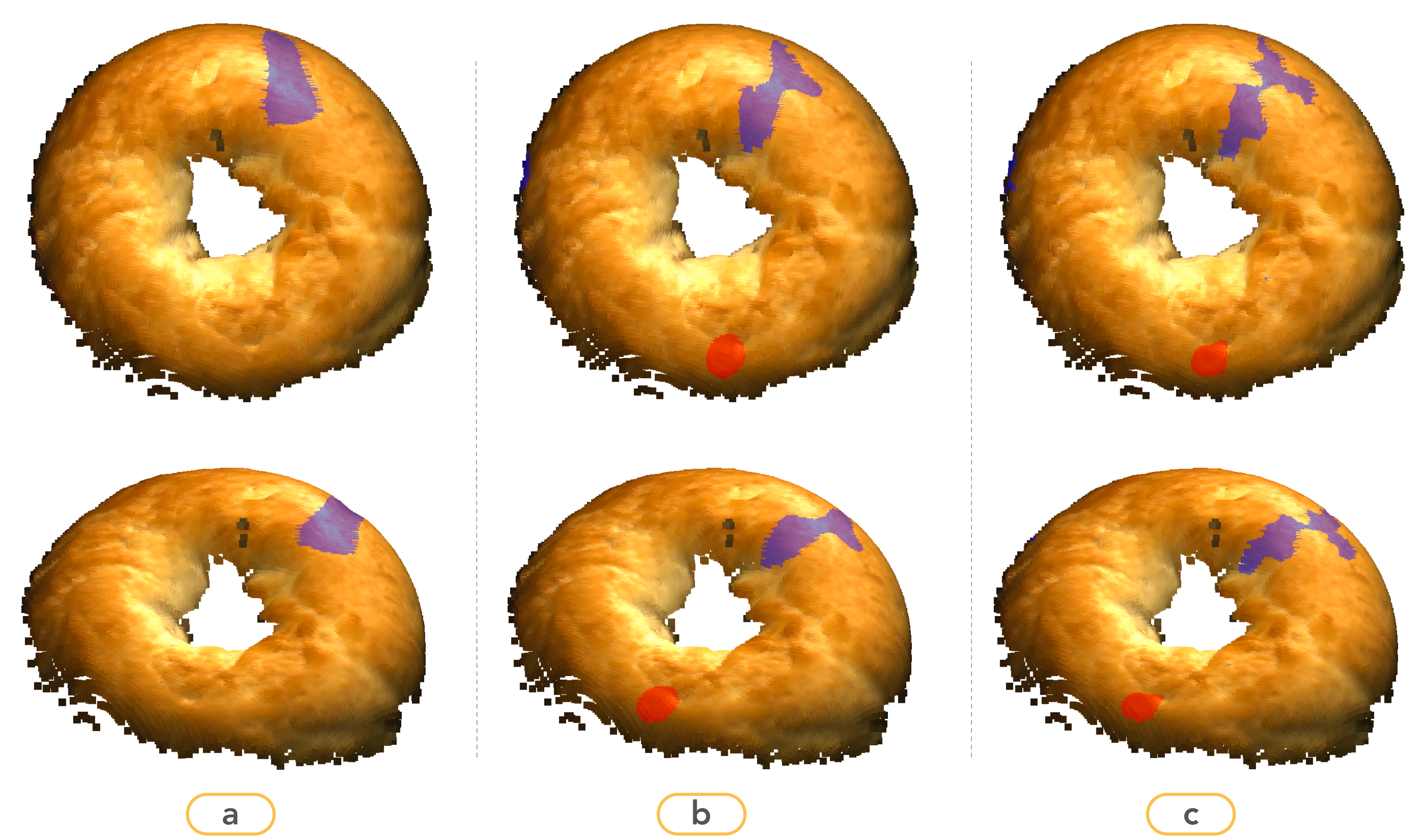}
    \caption{Visualization of the effect of varying the octaves ($o$) on the generated anomalies in a 3D point cloud: (a) $o=1$ produces smoother, single-scale anomalies; (b) $o=2$ adds an intermediate level of detail; and (c) $o=4$ results in more complex, multi-scale anomalies with both large-scale structures and fine-grained details.}
    \label{fig:octave}
\end{figure*}

Figures \ref{fig:noise_scale}, \ref{fig:perturb_strength}, and \ref{fig:octave} visualize the effects of varying these key parameters while keeping others constant. Through these visualizations, we demonstrate how our method enables fine-grained control over the generated anomalies. Noise scale ($s$) controls the spatial distribution and size of anomalies, perturbation strength ($\alpha$) determines the magnitude of deformations, and octaves ($o$) affect the complexity and detail level of the generated patterns. This illustrates how different parameter combinations can produce a diverse spectrum of anomaly patterns, from large pronounced defects to subtle surface variations.

\subsection{Cross-Category Consistency}
\begin{figure*}[ht]
    \centering
    \includegraphics[width=\linewidth]{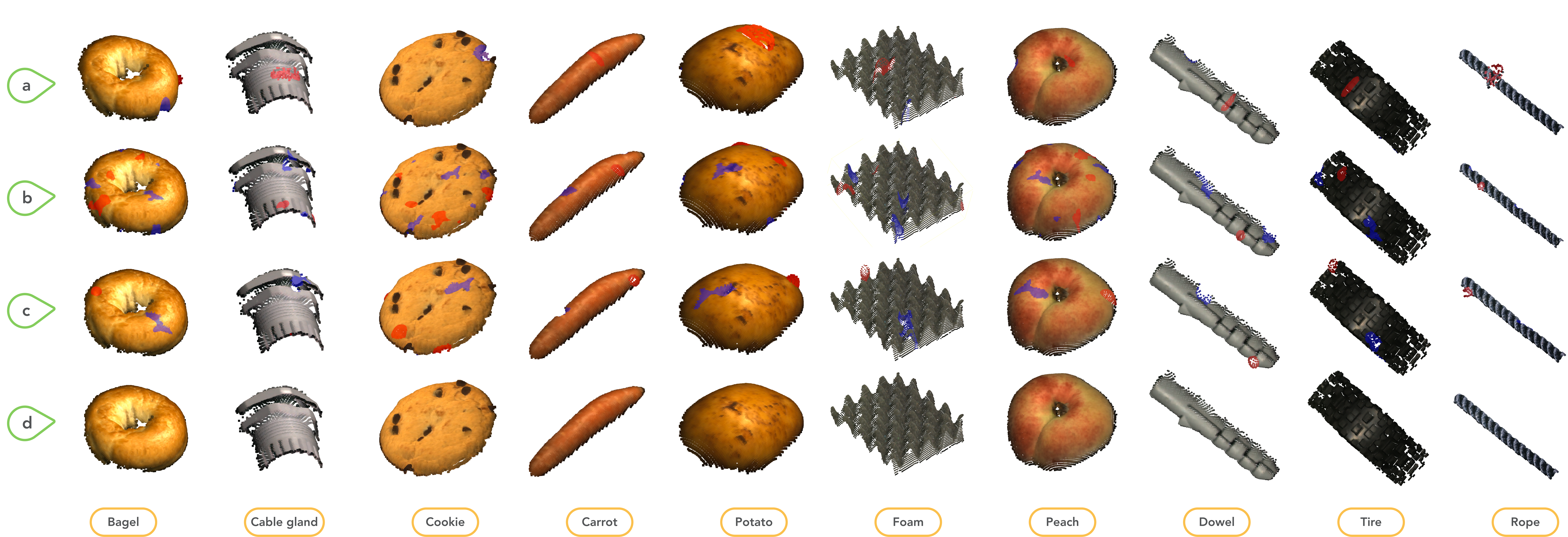}
    \caption{Visualization of our method's adaptability across various object categories from the MVTec 3D-AD dataset\cite{bergmannMVTec3DADDataset2022}. Each column represents a different object category (Bagel, Cable gland, Cookie, Carrot, Potato, Foam, Peach, Dowel, Tire, and Rope), while rows show different parameter profiles: (a) Pronounced defects with large deformations ($s=1.0$, $o=1$, $p=0.7$, $l=2.0$, $\tau=0.5$, $\rho=0.03$, $r=64$, $\alpha=0.1$); (b) Subtle surface variations with fine details ($s=3.0$, $o=3$, $p=0.4$, $l=2.0$, $\tau=0.6$, $\rho=0.08$, $r=64$, $\alpha=0.02$); (c) Medium deformations with balanced parameters ($s=2.0$, $o=2$, $p=0.5$, $l=2.0$, $\tau=0.6$, $\rho=0.05$, $r=64$, $\alpha=0.05$); and (d) Normal point clouds without anomalies for reference. Red points indicate protrusions, while blue points indicate intrusions. Our method successfully adapts to each object's unique geometry while maintaining consistent anomaly patterns.}
    \label{fig:cross_category}
\end{figure*}

To demonstrate the robustness of our method across different object geometries, we applied various anomaly parameter profiles to multiple object categories from the MVTec 3D-AD dataset. Figure \ref{fig:cross_category} shows a matrix visualization where each row represents a different object category, and each column shows a specific anomaly profile, ranging from pronounced defects to subtle variations, as well as the original normal point cloud for reference. We systematically tested three parameter profiles: (1) pronounced defects with large deformations (noise scale $s=1.0$, octaves $o=1$, persistence $p=0.7$, lacunarity $l=2.0$, threshold $\tau=0.5$, mask ratio $\rho=0.03$, grid resolution $r=64$, perturbation strength $\alpha=0.1$), (2) medium deformations with balanced parameters (noise scale $s=2.0$, octaves $o=2$, persistence $p=0.5$, lacunarity $l=2.0$, threshold $\tau=0.6$, mask ratio $\rho=0.05$, grid resolution $r=64$, perturbation strength $\alpha=0.05$), and (3) subtle surface variations with fine details (noise scale $s=3.0$, octaves $o=3$, persistence $p=0.4$, lacunarity $l=2.0$, threshold $\tau=0.6$, mask ratio $\rho=0.08$, grid resolution $r=64$, perturbation strength $\alpha=0.02$).

Results demonstrate that our method successfully adapts to the specific geometry of each object, producing consistent anomaly patterns while respecting the local surface characteristics.

\subsection{Parameter Tuning via Grid Search}
Building on the individual exploration of key parameters, we further investigated the coupling effects between noise scale ($s$) and octaves ($o$), as these parameters showed significant influence on anomaly characteristics as demonstrated in Figures \ref{fig:noise_scale} and \ref{fig:octave}. To systematically analyze their joint impact and facilitate the identification of optimal parameter settings for specific industrial anomaly synthesis tasks, we employed a grid search approach that visualizes the generated anomalies across combinations of parameter values.

\begin{figure*}[ht]
    \centering
    \includegraphics[width=\linewidth]{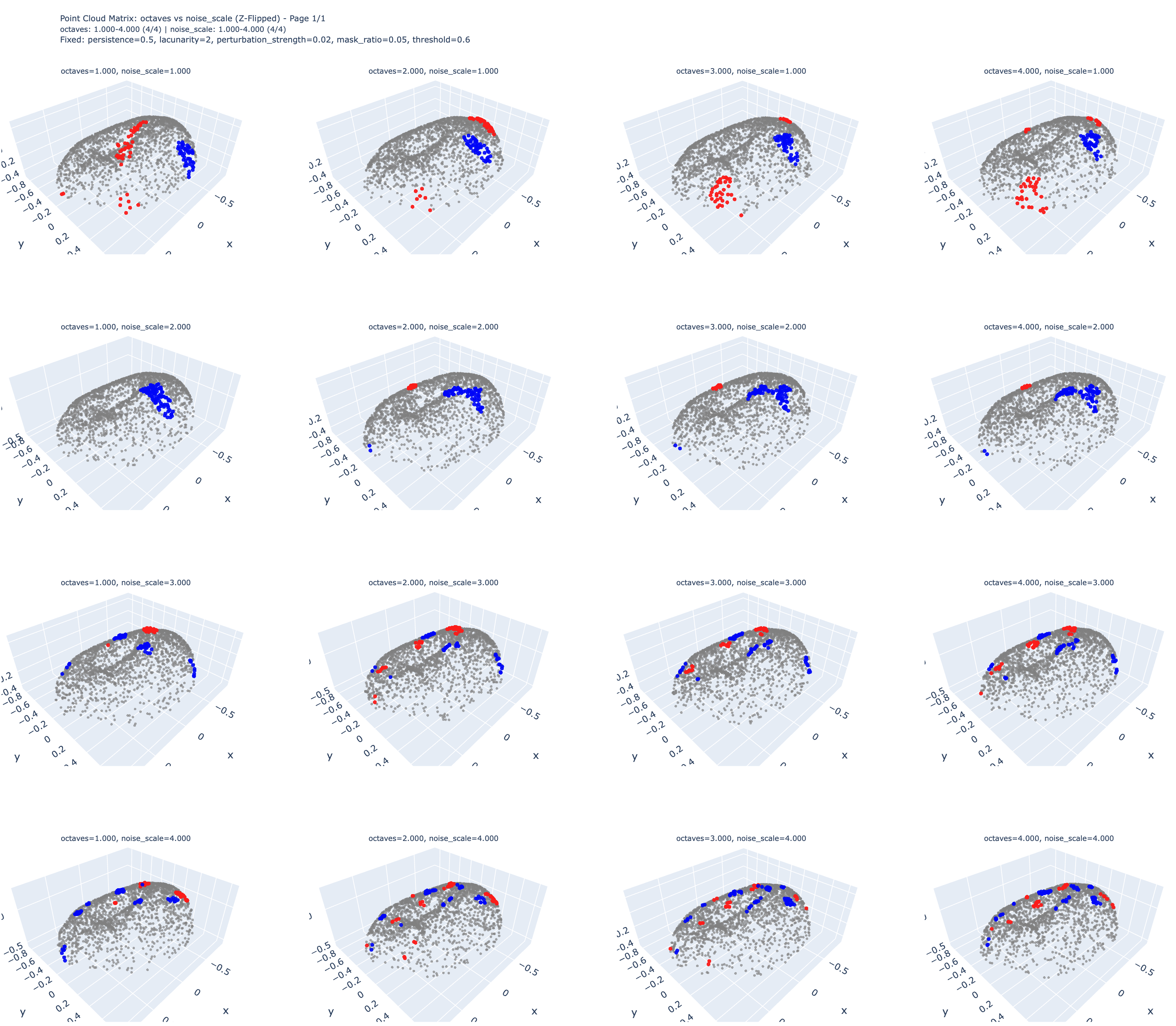}
    \caption{Grid search matrix visualization generated by our grid serach tool showing the coupling effects of noise scale ($s$) and octaves ($o$) on generated anomalies. Rows represent increasing values of noise scale $s \in \{1, 2, 3, 4\}$ from top to bottom, and columns represent increasing values of octaves $o \in \{1, 2, 3, 4\}$ from left to right. Red points indicate protrusions, while blue points indicate intrusions.}
    \label{fig:param_grid}
\end{figure*}

Figure \ref{fig:param_grid} presents a matrix visualization where noise scale ($s$) varies from 1.0 to 4.0 (rows) and octaves ($o$) ranges from 1 to 4 (columns), creating 16 unique parameter combinations. Other parameters were kept as constant at our established default values (persistence $p=0.5$, lacunarity $l=2.0$, threshold $\tau=0.6$, mask ratio $\rho=0.05$, perturbation strength $\alpha=0.02$, and grid resolution of 64) to isolate the effects of the studied parameters. Each plot in the matrix contains a 3D interactive visualization of the resulting anomaly pattern for that specific parameter combination, despite due to the limitations of the research paper format, we are unable to provide the interactive visualization here.

This systematic exploration reveals several key insights about parameter interactions. The combination of low noise scale ($s=1.0$) with high octaves ($o=4$) produces anomalies with large overall structures but with complex, fine-grained surface details. Conversely, high noise scale ($s=4.0$) with low octaves ($o=1$) generates multiple small, smooth anomalies distributed across the surface. The transition along diagonal elements of the matrix (from low $s$ /low $o$ to high $s$ /high $o$) shows a gradual progression from few large, simple anomalies to numerous small, detailed deformations.

These parameter coupling effects enable precise control over anomaly characteristics that would be difficult to achieve by adjusting individual parameters independently. To support further exploration and reproducible research, we have included the parameter grid search toolkit with GUI in our publicly available codebase, allowing researchers and practitioners to generate customized anomaly patterns tailored to specific industrial inspection needs.

\section{Conclusion}
Our work introduces a novel 3D anomaly synthesis method based on Perlin noise and surface parameterization, addressing the gap in 3D anomaly generation for industrial anomaly detection. Through comprehensive visualizations, we demonstrate that our method generates diverse and physically plausible anomalies. The flexibility of our approach, enabled by tunable noise parameters, allows for the generation of anomalies tailored to specific industrial needs. While our results are currently limited to visual demonstrations, they highlight the potential of our method for generating realistic synthetic data, which is crucial for adapting pretrained foundational models for the 3D anomaly detection task. Future work will focus on integrating our method with advanced 2D generative models for multimodal anomaly generation and validating its effectiveness on diverse 3D industrial datasets.

\bibliographystyle{IEEEtran}
\bibliography{main}

\end{document}